\documentclass[sigconf]{acmart}
\AtBeginDocument{%
  }

 \copyrightyear{2025} 
\acmYear{2025} 
\setcopyright{acmlicensed}\acmConference[ICMR '25]{Proceedings of the 2025
 International Conference on Multimedia Retrieval}{June 30-July 3,
 2025}{Chicago, IL, USA}
 \acmBooktitle{Proceedings of the 2025 International Conference on Multimedia
 Retrieval (ICMR '25), June 30-July 3, 2025, Chicago, IL, USA}
 \acmDOI{10.1145/3731715.3733305}
 \acmISBN{979-8-4007-1877-9/2025/06}

\begin{document}

\title{DGFNet: End-to-End Audio-Visual Source Separation Based on Dynamic Gating Fusion}


\author{Yinfeng Yu}
\affiliation{%
  \department{School of Computer Science and Technology}
  \institution{Xinjiang University} 
  \city{Urumqi}
  \country{China}
}
\email{yuyinfeng@xju.edu.cn}

\author{Shiyu Sun}
\authornote{Corresponding author.}
\affiliation{%
 \department{School of Computer Science and Technology}
  \institution{Xinjiang University}
  \city{Urumqi}
  \country{China}
}
\email{107552304141@stu.xju.edu.cn}
\renewcommand{\shortauthors}{Yinfeng Yu and Shiyu Sun}

\begin{abstract}
  Current Audio-Visual Source Separation methods primarily adopt two design strategies. The first strategy involves fusing audio and visual features at the bottleneck layer of the encoder, followed by processing the fused features through the decoder. However, when there is a significant disparity between the two modalities, this approach may lead to the loss of critical information. The second strategy avoids direct fusion and instead relies on the decoder to handle the interaction between audio and visual features. Nonetheless, if the encoder fails to integrate information across modalities adequately, the decoder may be unable to effectively capture the complex relationships between them. To address these issues, this paper proposes a dynamic fusion method based on a gating mechanism that dynamically adjusts the modality fusion degree. This approach mitigates the limitations of solely relying on the decoder and facilitates efficient collaboration between audio and visual features. Additionally, an audio attention module is introduced to enhance the expressive capacity of audio features, thereby further improving model performance. Experimental results demonstrate that our method achieves significant performance improvements on two benchmark datasets, validating its effectiveness and advantages in Audio-Visual Source Separation tasks.
\end{abstract}


\begin{teaserfigure}
  \includegraphics[width=\textwidth]{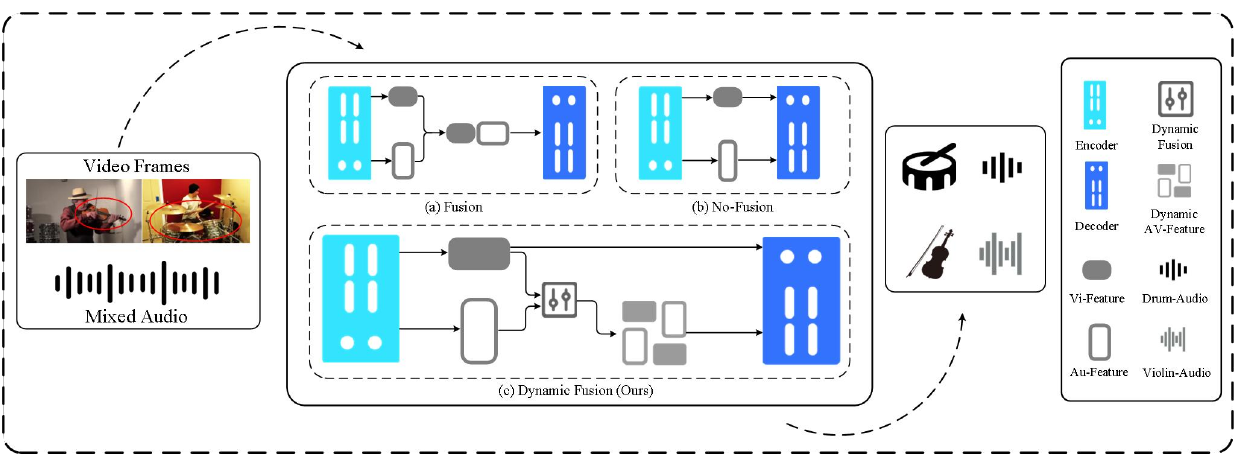}
  \caption{Comparison between previous mainstream methods (a), (b), and the proposed Dynamic Fusion (c).}
  \Description{The design approach is described and compared with the other two approaches.}
  \label{fig:head-figure}
\end{teaserfigure}

\maketitle

\section{Introduction}
Audio-Visual learning \cite{DBLP:conf/mm/ChengWPFZ20,DBLP:conf/eccv/AfourasOCZ20} involves integrating information from both visual and auditory modalities to address real-world problems. In a noisy, multi-source environment, we can effortlessly utilize visual scene information to filter out unwanted noise and identify the sounds we desire. This is the problem of Audio-Visual Source Separation, also known as the "Cocktail Party Problem" \cite{DBLP:journals/neco/HaykinC05}. While source separation focuses on isolating target audio signals, Audio-Visual learning has also been extensively explored in other tasks such as sound-guided navigation and acoustic perception, where agents use multimodal information to localize, interpret, or follow sound sources in complex environments \cite{SAAVN,fsaavn,YinfengIJCAI2023MACMA,ttt}. This paper focuses on sound source separation to identify and segregate different sound components within a given audio mixture.

To create a multi-source environment, we follow the "Mix and Separate" framework \cite{DBLP:conf/icassp/HersheyCRW16,DBLP:conf/icassp/YuKT017} by mixing multiple audio signals to generate artificially complex auditory representations, which are then used as a self-supervised task to separate individual sounds from the mixtures. Early works \cite{DBLP:journals/tog/EphratMLDWHFR18,Zhao_2018_ECCV,DBLP:conf/iccv/GaoG19} that incorporated visual information into the audio source separation problem have demonstrated the feasibility of guiding audio source separation through visual cues.

Currently, Audio-Visual Source Separation commonly employs an encoder-decoder architecture, where the encoder separately extracts visual and audio features, fuses these features at the bottleneck layer, and subsequently, the decoder generates spectrogram masks for separating audio. However, many methods \cite{Zhao_2018_ECCV,DBLP:conf/iccv/GaoG19,DBLP:conf/iccv/ZhaoGM019,DBLP:conf/cvpr/GanHZT020,DBLP:conf/accv/0001R20,DBLP:journals/tmm/JiMXLS23,DBLP:conf/wacv/IslamNKWYT24} utilize channel-wise concatenation or pixel-wise multiplication for feature fusion. This approach may result in the loss or confusion of critical information, especially when there is a significant disparity between the modalities. The multimodal feature learning and audio decoding processes can become confused, leading to separated sounds that are either missing or overlapped with other audio sources. To address this issue, \cite{DBLP:conf/cvpr/GaoG21} enforced cross-modal consistency by setting appropriate loss functions, \cite{DBLP:conf/iclr/ChengLW023} designed a Filter-Recovery network to enhance the separated audio, and \cite{DBLP:conf/accv/0001R20,DBLP:journals/tmm/JiMXLS23} employed multi-stage separation to compensate for initial audio separation deficiencies. Recently, \cite{DBLP:conf/cvpr/ChenZLYZS23} adopted the second design strategy by handling the interaction between the two modalities within the decoder instead of fusing them at the bottleneck layer, thereby making feature learning and audio decoding processes more straightforward. However, if the encoder fails to integrate information across modalities sufficiently, the decoder may still be unable to effectively capture the complex relationships between the modalities.

We propose a Dynamic Gating Fusion Network (DGFNet) for End-to-End Audio-Visual Source Separation to address the abovementioned limitations. The core idea of this approach lies in dynamically learning the importance of different modality features, enabling more refined feature adjustment. During the fusion process, the model computes the gating weights between the audio features and the fused Audio-Visual features at the bottleneck layer, generating weight coefficients through an activation function. These coefficients determine the contribution ratio of visual and audio features during the fusion process, thereby retaining the most task-relevant information in the final fused features. The model can dynamically adjust the modality fusion strategy based on the input data. Ultimately, the fused features replace the original audio, becoming an audio representation with integrated multimodal information rather than a single-modal audio feature. This multimodal fusion method significantly improves the quality of audio separation, particularly in complex background noise or multi-source scenarios, where it is more effective in identifying and separating target audio. Figure \ref{fig:head-figure} illustrates and compares our design concept with two other approaches.

The main contributions of this paper are as follows:
\begin{itemize}
	\item To the best of our knowledge, we are the first to propose the combination of bottleneck feature fusion with decoder interaction decoding strategy for the Audio-Visual Source Separation task.
	\item An audio attention module is introduced to enhance the expressive capability of audio features, further improving the model's performance.
	\item Through comparative experiments and ablation studies, we validate the effectiveness of the core design—dynamic gating fusion—and demonstrate significant performance improvements in Audio-Visual Source Separation tasks across two benchmark datasets.
\end{itemize}

\section{Related Work}

\subsection{Audio-Visual Source Separation}

Source separation is a classic problem in the field of audio signal processing. In the early stages, researchers predominantly relied on Independent Component Analysis (ICA) \cite{Hussain} or Non-negative Matrix Factorization (NMF) \cite{DBLP:conf/eccv/GaoFG18} for source separation. With the development of deep learning, significant advancements in source separation technologies have been achieved by cleverly integrating visual information with audio signals. Recent work on Audio-Visual Source Separation can be roughly categorized into speaker separation \cite{DBLP:conf/cvpr/GaoG21,DBLP:conf/asru/PanWMGKHR23,pegg2024rtfsnet,DBLP:conf/iclr/ChengLW023}, instrument separation \cite{DBLP:conf/cvpr/ChenZLYZS23,DBLP:journals/corr/abs-2110-13412,DBLP:conf/wacv/YeYT24}, natural sound source separation \cite{DBLP:conf/eccv/TzinisWRH22,Efthymios}, and active Audio-Visual Source Separation \cite{DBLP:conf/iccv/MajumderAG21,DBLP:conf/eccv/MajumderG22}. For speaker separation, \cite{pegg2024rtfsnet} adopts a unique time-frequency domain algorithm combined with an attention-based information fusion technique and a new spectral mask separation method, significantly reducing model complexity while maintaining performance. For instruments, \cite{DBLP:conf/cvpr/ChenZLYZS23} redefines the Audio-Visual separation task through a flexible query expansion mechanism, ensuring cross-modal consistency and instrument separation. For natural sound sources, \cite{DBLP:conf/eccv/TzinisWRH22} employs a more efficient self-attention mechanism, including Audio-Visual self-attention and cross-modal attention, to capture dependencies between audio and video. For active Audio-Visual Source Separation, \cite{DBLP:conf/iccv/MajumderAG21} utilizes visual and auditory cues and a new reward mechanism to guide the agent’s movement in the environment, improving the quality of audio predictions.

\subsection{Vision Transformer}

Inspired by the significant achievements of the Transformer model in the field of natural language processing \cite{Vaswani}, this architecture was initially introduced into the computer vision field, aiming to make a breakthrough in image classification tasks, such as through the implementation of Vision Transformer (ViT) \cite{DBLP:conf/iclr/DosovitskiyB0WZ21}. Given the Transformer’s outstanding ability to capture long-range dependencies, subsequent research \cite{Mao,Yuan} continuously optimized and enhanced ViT, pushing its performance limits and enabling it to outperform traditional Convolutional Neural Networks (CNNs) in various scenarios. As research in the Audio-Visual Source Separation field deepened, researchers turned their attention to Transformers, leading to the emergence of several notable works \cite{DBLP:conf/cvpr/ChenZLYZS23,DBLP:journals/corr/abs-2110-13412,DBLP:conf/icassp/Kalkhorani00XW24,DBLP:conf/eccv/TzinisWRH22,Efthymios}.

\begin{figure*}[h]
	\centering
	\includegraphics[width=\linewidth]{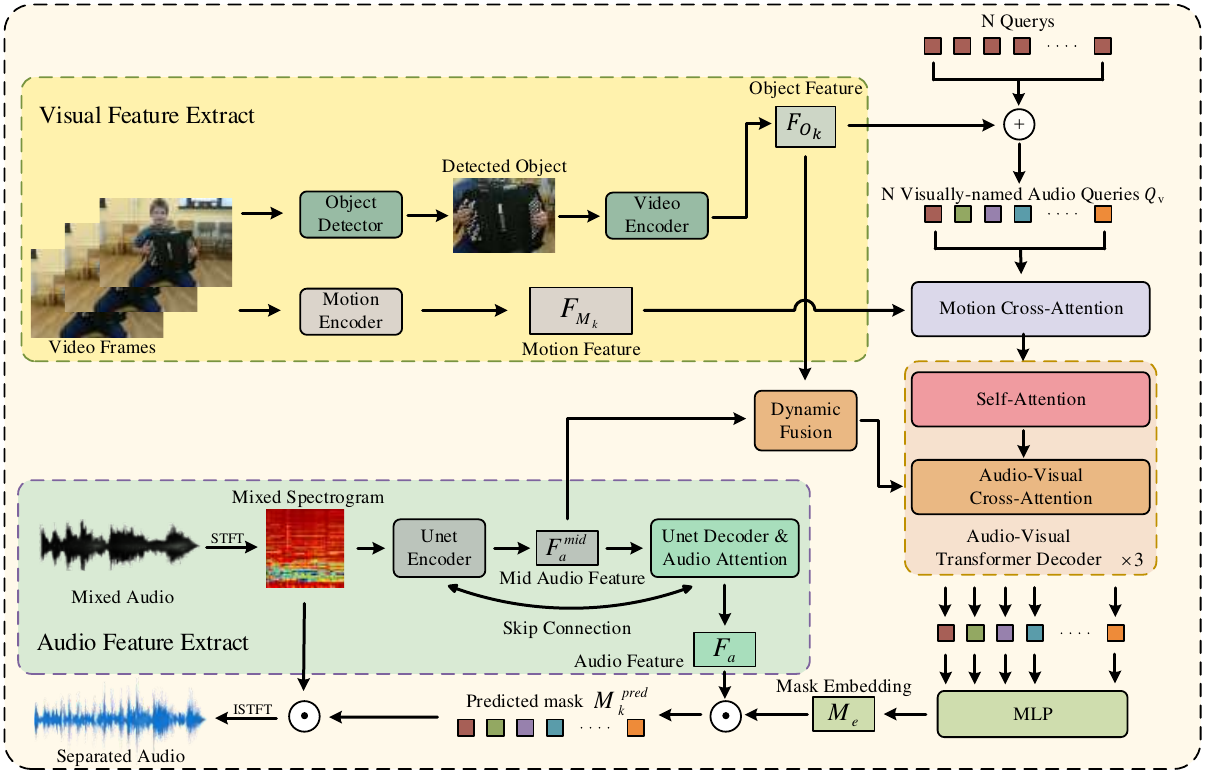}
	\caption{The overall architecture of the DGFNet. The model comprises the Audio-Visual feature extraction network, Audio-Visual feature fusion network, and query-based Audio-Visual Transformer network. The feature extraction network extracts object and motion features from video segments $V_{k}$ using an object detector, motion encoder, and image encoder. The audio feature extraction network extracts features from the mixed audio signal $s_{mix}(t)$. The Audio-Visual feature fusion network integrates visual and audio features through the Dynamic Gating Fusion Module (DGFM). Finally, the Audio-Visual Transformer network generates mask embeddings using the query mechanism for target audio separation.}
        \Description{The architecture of our network is presented.}
	\label{fig:Pipeline of our method}
\end{figure*}

\section{Proposed Method}

We introduce an Audio-Visual Source Separation method that dynamically integrates visual and audio features using a gating mechanism. In Section 3.1, we provide a brief overview of our model. Then, in Section 3.2, we describe the Audio-Visual feature extraction network, in Section 3.3, we present our Dynamic Gating Fusion Module (DGFM), and in Section 3.4, we introduce the Audio-Visual Transformer module with a query mechanism that we utilize.

\subsection{Overview}

Our work draws inspiration from the "mixing and separation" framework proposed in \cite{Zhao_2018_ECCV}. Given $\mathit{K}$ video segments containing audio signals $\{ V_{k}, s_{k}(t)\}_{k \in [1, K]}$, we first mix the audio signals to obtain the mixed audio signal $s_{mix}(t)=\sum _{k=1}^{K}s_{k}(t)$. Our goal is to effectively separate each audio signal $s_{k}(t)$ from the mixed audio signal $s_{mix}(t)$ by utilizing the visual information contained in each video segment $V_{k}\in \mathbb{R}^{3\times T_{k}\times H_{k}\times W_{k}}$. The architecture of our model is shown in Figure \ref{fig:Pipeline of our method}.

\subsection{Audio-Visual Feature Extraction}

\subsubsection{Visual NetWork}

The visual network is required to extract object-level visual features and motion features. For visual features, following the work of \cite{DBLP:conf/iccv/GaoG19}, we adopt a pre-trained Faster R-CNN \cite{DBLP:journals/pami/RenHG017} object detector to detect objects $O_{k}$ in the input video $V_{k}$. Similarly to the approach used in \cite{DBLP:conf/cvpr/TianHX21}, we use a pre-trained ResNet-18, followed by a linear layer and max pooling to generate object-level features $F_{O_{k}}\in\mathbb{R}^{C_{O}}$. For motion features, inspired by previous works \cite{DBLP:conf/iccv/ZhaoGM019,DBLP:conf/cvpr/GanHZT020,Zhu_Rahtu_2022,DBLP:journals/corr/abs-2110-13412,DBLP:conf/cvpr/ChenZLYZS23}, we input the video frames into a pre-trained I3D video encoder from FAME \cite{DBLP:conf/cvpr/DingLYQXCWX22}, and after spatial pooling, we obtain the motion features $F_{M_{k}}\in \mathbb{R}^{C_{M}\times T'_{k}}$.

\subsubsection{Audio NetWork}

The audio network adopts the U-Net \cite{DBLP:conf/miccai/RonnebergerFB15} style network used in \cite{Zhao_2018_ECCV,DBLP:conf/iccv/GaoG19,DBLP:journals/corr/abs-2110-13412}, with skip connections between the encoder and decoder. The input to the network is the spectrogram $S_{mix}\in \mathbb{R}^{F\times T}$, which is obtained by applying Short-Time Fourier Transform (STFT) \cite{DBLP:conf/icassp/GriffinL83} to the mixed audio signal $s_{mix}(t)$. We adopt a similar feature extraction process as in \cite{DBLP:conf/cvpr/ChenZLYZS23}, adding an audio attention module after each upsampling layer in the decoder. As a result, the intermediate audio features $F_{a}^{mid}\in \mathbb{R}^{C_{A}\times \frac{F}{S}\times \frac{T}{S}}$ are obtained during the upsampling process, and the final audio features $F_{a}\in \mathbb{R}^{C'_{A}\times F\times T}$ are obtained after the final upsampling layer.

\subsubsection{Audio Attention Module}

\begin{figure}
	\centering
	\includegraphics[width=7cm]{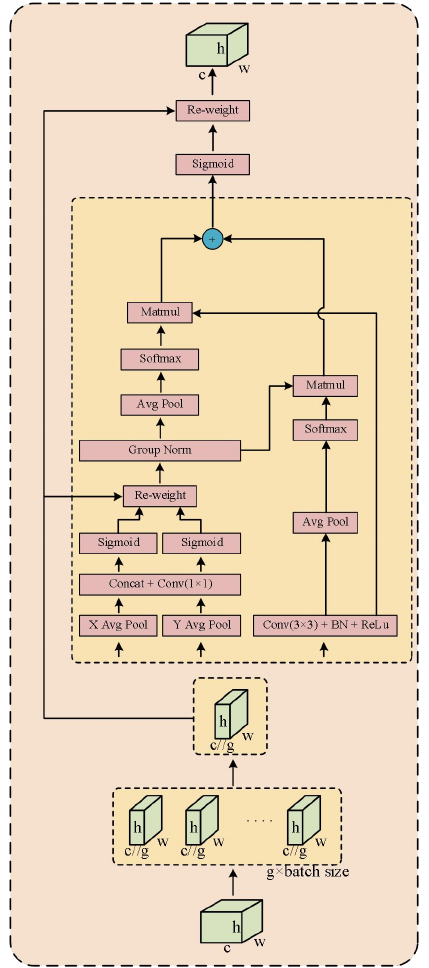}
	\caption{Audio Attention Module.}
    \Description{The architecture of our audio attention module is presented.}
	\label{fig:audio_attention}
\end{figure}

Our Audio Attention module is inspired by the Efficient Multi-scale Attention module (EMA) proposed in \cite{DBLP:conf/icassp/OuyangHZLGZH23}. The EMA uses the shared 1×1 convolutional branch from the Revisit Coordinate Attention (CA) module and names it the 1×1 branch. Additionally, to aggregate multi-scale spatial structural information, EMA places a 3×3 convolution kernel in parallel with the 1×1 branch, referred to as the 3×3 branch. Considering feature grouping and multi-scale structures, EMA effectively establishes short- and long-range dependencies, generating better pixel-level attention for high-level feature maps, thus improving performance. We replace the single 3×3 convolution with a combination of convolution, batch normalization, and ReLU activation functions to enhance training stability, accelerate convergence, and improve expressive capability. The structure of the Audio Attention module is shown in Figure \ref{fig:audio_attention}.

\subsection{Dynamic Gating Fusion Module}

\begin{figure}[ht]
	\centering
	\includegraphics[width=7cm]{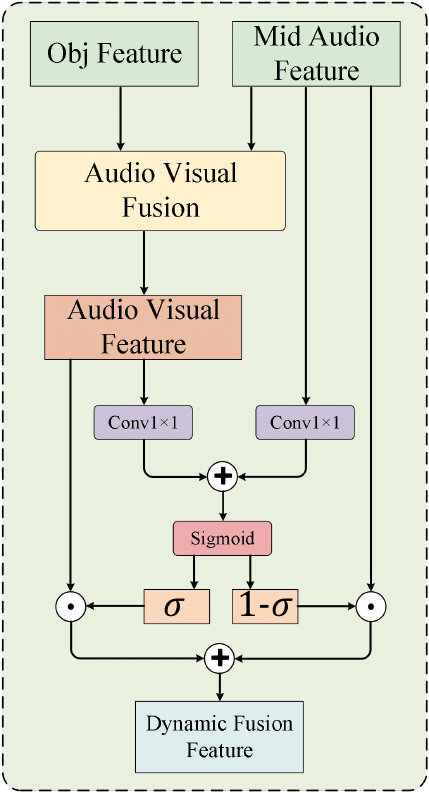}
	\caption{Dynamic Gating Fusion Module.}
    \Description{The architecture of our DGFM is presented.}
	\label{fig:DGFM}
\end{figure}

Inspired by the approach of processing Audio-Visual features at the bottleneck of the encoder-decoder in \cite{DBLP:conf/iccv/GaoG19,DBLP:conf/cvpr/ChenZLYZS23}, we design the Dynamic Gating Fusion Module (DGFM), as illustrated in Figure~\ref{fig:DGFM}. The module takes object features $F_{O_{k}}$ and intermediate audio features $F_{a}^{mid}$ as input. First, the two features are fused by pixel-wise multiplication along the channel dimension, resulting in $F_{av}$. Then, both $F_{av}$ and $F_{a}^{mid}$ are passed through a 1×1 convolutional layer to generate two weight matrices. These matrices are summed and passed through a Sigmoid function to obtain the gating coefficient $\sigma$. The final fused feature is computed as $\sigma \cdot F_{av} + (1 - \sigma) \cdot F_{a}^{mid}$, and the resulting feature $F_d \in \mathbb{R}^{C_{A} \times \frac{F}{S} \times \frac{T}{S}}$ replaces $F_{a}^{mid}$ in the network.

From a theoretical perspective, this design enables data-dependent dynamic reweighting across modalities. The learnable gating coefficient $\sigma$ allows the model to adaptively modulate the contribution of audio and visual features based on the contextual input. When visual cues are clear, such as the presence of well-defined instrument appearances, a higher $\sigma$ increases the influence of visual features, providing stronger guidance for separation. Conversely, in the presence of occlusion or ambiguous visuals, the model reduces $\sigma$ to emphasize the audio modality. This adaptive gating mechanism improves the model’s robustness by mitigating over-reliance on noisy modalities and enhancing the discriminative power of the fused representation. It serves as a key component in enabling the model to generalize effectively across diverse and complex real-world Audio-Visual scenarios.

\subsection{Audio-Visual Transformer}

Our Audio-Visual Transformer decoder still adopts the architecture and query mechanism proposed in \cite{DBLP:conf/cvpr/ChenZLYZS23}. This module contains $N$ queries, and it uses the object features $F_{O_{k}}$, motion features $F_{M_{k}}$, and fused features $F_{d}$ to generate $N$ mask embeddings $M_{e}\in \mathbb{R}^{C_{M}\times N}$, where $N$ represents the predefined maximum number of instrument types.

\subsubsection{Audio Query}

The audio queries are represented as $Q\in \mathbb{R}^{C_{Q}\times N}$, which corresponds to different types of instruments. The object features $F_{O_{k}}$ are assigned to the corresponding queries in $Q$, and element-wise addition is performed to generate the "visual naming" queries $Q_{v}$. These queries are then fed into the cross-attention layer of the Transformer decoder.

\subsubsection{Cross-modal Interaction Layers}

The query $Q_{v}$ first interacts with the motion features $F_{M_{k}}$ along the time dimension in the motion cross-modal attention layer. Then, a Feed-Forward Network (FFN) is applied to generate the motion decoding queries $Q'$. These queries $Q'$ are then input into three Audio-Visual cross-modal attention decoder layers to adaptively interact with the features $F_{d}$. Each decoder layer comprises self-attention and multi-head attention, generating $N$ audio separation embeddings $A_{e}$. The specific process is as follows:
\begin{equation}
	Q' = FFN(\text{Motion-Attention}(Q_{v}, F_{M_{k}}, F_{M_{k}})),
\end{equation}
\begin{equation}
	F'_{d} = \text{Self-Attention}(F_{d}, F_{d}, F_{d}),
\end{equation}
\begin{equation}
	Q'' = \text{Self-Attention}(Q', Q', Q'),
\end{equation}
\begin{equation}
	A_{e} = \text{AV-Attention}(Q'', F'_{d}, F'_{d}).
\end{equation}

\subsubsection{Mask Prediction}

The $N$ audio separation embeddings $A_{e}$ obtained from the decoder layers are input into an MLP, which outputs $N$ mask embeddings $M_{e}$. Then, the corresponding mask embeddings in $M_{e}$ are multiplied element-wise with the audio features $F_{a}$ from the U-Net decoder to obtain the predicted mask map $M^{pred}_{k}\in \mathbb{R}^{F\times T}$for the separated audio spectrogram. Finally, the mixed spectrogram $S_{mix}$ is multiplied by the predicted mask $M^{pred}_{k}$ to obtain the separated audio spectrogram $S_{k}$. The specific formula is as follows:
\begin{equation}
	S_{k} = S_{mix} \odot M_{k}^{pred}.
\end{equation}
Finally, the separated audio spectrogram $S_{k}$ is transformed back into the separated audio signal $s_{k}(t)$ by applying the inverse Short-Time Fourier Transform (iSTFT).

\subsubsection{Training objective}

Based on the work of \cite{Zhao_2018_ECCV,DBLP:conf/cvpr/ChenZLYZS23}, we optimize the generated spectrogram masks. The formula for computing the ground truth mask, denoted as $M_{k}^{GT}$, is as follows:
\begin{equation}
	M_{k}^{GT}(t, f) = \frac{S_{k}(t, f)}{S_{mix}(t, f)}.
\end{equation}
$(t, f)$ represents the time-frequency coordinates, and $k$ denotes the $k$-th video. We optimize the model by optimizing a pixel-wise $L1$ Loss \cite{Zhao_Gallo_Frosio_Kautz_2017}. The final loss function $L_{sep}$ is given by:
\begin{equation}
	L_{sep} = \sum_{k=1}^{K} ||M_{k} - M_{k}^{GT}||_{1}.
\end{equation}
$K$ denotes the number of mixed sound sources in $S_{mix}$.

\section{Experiments}

\subsection{Experimental Settings}

\subsubsection{Datasets}

We conducted experiments on two widely used datasets: MUSIC \cite{Zhao_2018_ECCV} and MUSIC-21 \cite{DBLP:conf/iccv/ZhaoGM019}. The MUSIC dataset contains 685 untrimmed music solo and duet videos from 11 different instrument categories. The MUSIC-21 dataset expands the MUSIC dataset, consisting of 1365 untrimmed music solo and duet videos from 21 instrument categories. 

\subsubsection{Evaluation metrics}

We used three common evaluation metrics: Signal-to-Distortion Ratio (SDR), Signal-to-Interference Ratio (SIR), and Signal-to-Artifacts Ratio (SAR), and evaluated the results using the mir\textunderscore eval library \cite{Raffel_McFee}. Higher values for all metrics indicate better performance.

\subsubsection{Baselines}

We compared our method with several representative works from recent years. Due to the missing portions of videos in the MUSIC and MUSIC-21 datasets available on the internet, we used the results reported in \cite{DBLP:conf/cvpr/ChenZLYZS23} for some of the works. For the MUSIC dataset, the results of The Sound of Pixels \cite{Zhao_2018_ECCV}, Co-Separation \cite{DBLP:conf/iccv/GaoG19}, and CCoL \cite{DBLP:conf/cvpr/TianHX21} are taken from \cite{DBLP:conf/cvpr/ChenZLYZS23}. For the AVPC-RCoP \cite{10180219} dataset, we used the results reported by the authors. Lastly, we used iQuery \cite{DBLP:conf/cvpr/ChenZLYZS23} as the baseline and compared the results using the same training settings in our experimental environment.

On the MUSIC-21 dataset, we compared our results with The Sound of Pixels \cite{Zhao_2018_ECCV}, Co-Separation \cite{DBLP:conf/iccv/GaoG19}, The Sound of Motions \cite{DBLP:conf/iccv/ZhaoGM019}, AVPC-RCoP \cite{10180219}, and iQuery \cite{DBLP:conf/cvpr/ChenZLYZS23}. The results for \cite{Zhao_2018_ECCV, DBLP:conf/iccv/GaoG19, DBLP:conf/cvpr/GanHZT020, DBLP:conf/iccv/ZhaoGM019} are all taken from \cite{DBLP:conf/cvpr/ChenZLYZS23}. For \cite{10180219}, we used their reported results. Similarly, we used iQuery \cite{DBLP:conf/cvpr/ChenZLYZS23} as the baseline and compared the results using the same training settings in our experimental environment.

\begin{figure*}[htbp]
	\centering
	\includegraphics[width=\linewidth]{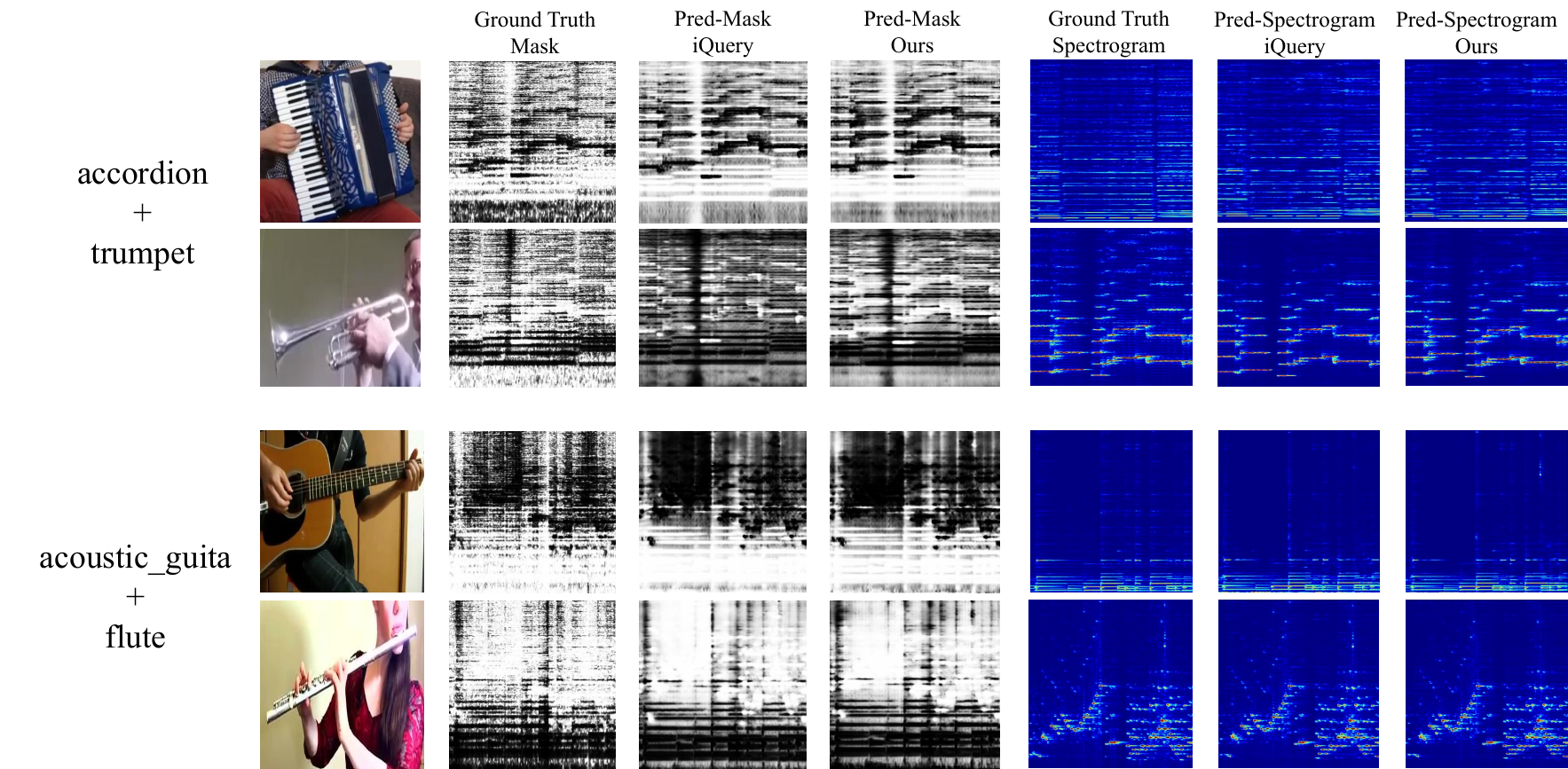}
	\caption{Qualitative results on MUSIC test set.}
    \Description{The performance differences between our method and others are compared through visualization.}
	\label{fig:qualitative}
\end{figure*}

\subsubsection{Implementation Details}

We downsample the audio to 11kHz, each audio clip lasting approximately 6 seconds. We perform Short-Time Fourier Transform (STFT) on the audio using a 1022-point Hann window and a hop size 256, resulting in a 512×256 time-frequency representation. The spectrogram is then resampled on a logarithmic frequency scale, yielding a spectrogram with dimensions $T, F$ = 256. In each video frame, the detected objects are resized to 256×256 pixels and randomly cropped to a size of 224×224. The video frame rate is set to 1 FPS, and three frames are randomly selected as input to the object detection network. These settings are consistent with \cite{DBLP:conf/cvpr/ChenZLYZS23}.

For the MUSIC dataset, we use the pre-trained Faster R-CNN object detector from \cite{DBLP:conf/iccv/GaoG19} for object detection. For the MUSIC-21 dataset, we use the pre-trained Detic \cite{zhou2022detecting} object detector.

We use the AdamW optimizer with a decay weight of $10^{-4}$ for the Audio-Visual Transformer network and the Adam optimizer for other networks. The epoch is set to 100, and the batch size is set to 20. The remaining settings are consistent with\cite{DBLP:conf/cvpr/ChenZLYZS23}. The training is conducted on 3 NVIDIA Tesla V100 GPUs.

\subsection{Results}

\subsubsection{Quantitative evaluation}

Table \ref{tab:music} presents the quantitative results of sound separation using state-of-the-art methods on the MUSIC dataset. Regarding the SDR score, our method outperforms baseline model \cite{DBLP:conf/cvpr/ChenZLYZS23} by 0.62 dB. The SIR score is 0.27 higher, and the SAR score is 0.68 higher. For the quantitative results on the MUSIC-21 dataset, we provide a performance comparison in Table \ref{tab:music21}. Similarly, our method outperforms the baseline model regarding SDR, SIR, and SAR metrics. The performance on the first two datasets demonstrates our model's ability to separate musical sounds.

\subsubsection{Qualitative evaluation}

Figure \ref{fig:qualitative} presents a qualitative comparison between our proposed model and the iQuery model on the MUSIC test set. The first column of each category displays the ground truth spectrogram, followed by the predicted results from the iQuery model and our model. The comparison in the figure includes mask-based predictions (on the left) and spectrogram-based predictions (on the right), highlighting the differences in performance between the two methods in audio signal separation. Our model demonstrates a clearer spectrogram structure, better preserving the audio details during the separation process.

\subsubsection{Ablations of Fusion method}

Table \ref{tab:ab} shows the impact of different fusion methods on model performance. The Baseline \cite{DBLP:conf/cvpr/ChenZLYZS23} refers to the model without fusion at the bottleneck layer. "+Mul" indicates that, in addition to the baseline model, a fusion method has been incorporated where the visual and audio features are multiplied element-wise. +DGFM" means the addition of the dynamic gating fusion module only, which confirms the effectiveness of our dynamic fusion module.

\begin{figure}[t]
	\centering
	\includegraphics[width=\linewidth]{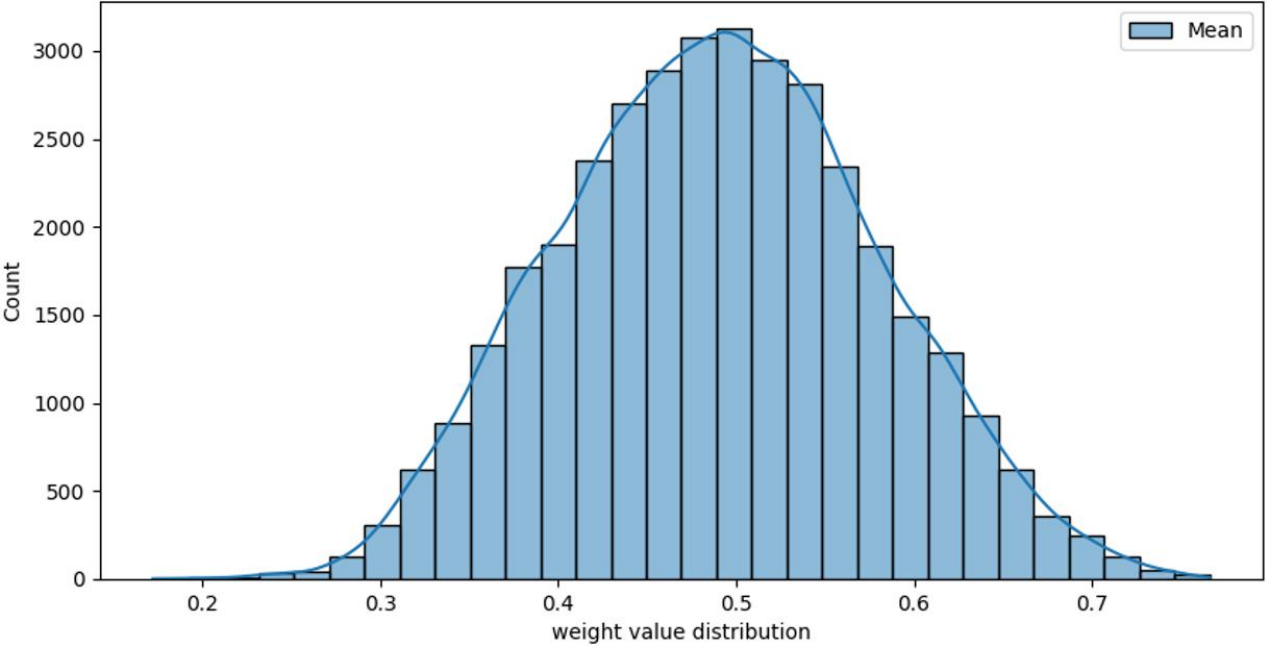}
	\caption{Distribution of Weight Coefficient Mean.}
	\label{fig:mean}
        \Description{Distribution of Weight Coefficient Mean.}
\end{figure}

\begin{figure}[htpb]
	\centering
	\includegraphics[width=\linewidth]{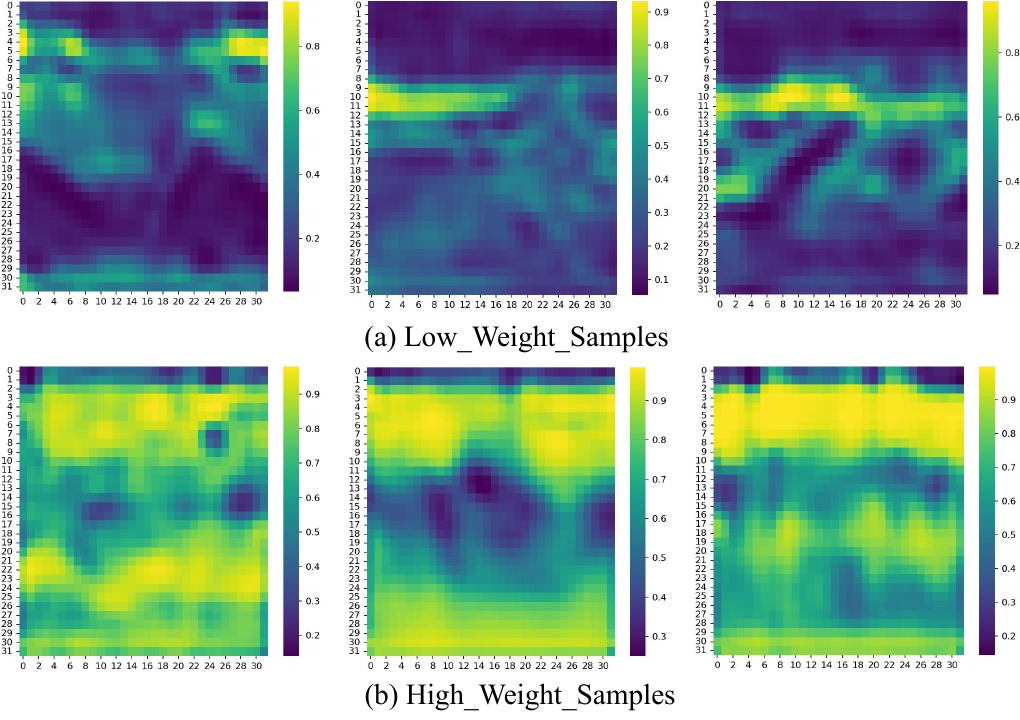}
	\caption{Comparison of High and Low Weight Sample Heatmaps.}
	\label{fig:heat}
        \Description{Heatmap Comparison of High and Low Weight Coefficient in Spatial Distribution.}
\end{figure}

\subsubsection{Effectiveness Analysis of the Dynamic Gating Fusion}

Figure \ref{fig:mean} illustrates the mean distribution of the weighting coefficient $\sigma$. It can be observed that the mean values of $\sigma$ exhibit a shape resembling a normal distribution, with most samples concentrated within the range of 0.4 to 0.6. This indicates that the dynamic weighting fusion mechanism effectively balances the integration of visual and audio features in most cases. Meanwhile, the number of samples at the distribution's extremes (below 0.3 and above 0.7) is relatively small, suggesting that the dynamic fusion mechanism can adaptively adjust the weighting to emphasize a particular modality in special cases. To further investigate this phenomenon, we compared the heat maps of samples at both distribution extremes, as shown in Figure \ref{fig:heat}. In the heatmap of low-weight samples (below 0.3), the purple regions (indicating lower weights) are dominant. This suggests that the model adopts a more conservative approach to feature fusion by reducing the contribution of visual features. Conversely, in the heatmap of high-weight samples (above 0.7), the yellow regions (indicating higher weights) predominantly cover a large portion of the space, demonstrating that the model relies more on visual features in these cases.

\begin{table}
	\caption{Audio-Visual sound separation results on MUSIC. Best results in \textbf{bold}.}
	\label{tab:music}
	\begin{tabular}{cccc}
		\toprule
		Methods&SDR$\uparrow$&SIR$\uparrow$&SAR$\uparrow$\\
		\midrule
		The Sound of Pixels&4.23&9.39&9.85\\
		Co-Separation&6.54&11.37&9.46\\
		CCol&7.74&13.22&11.54\\
		AVPC-RCoP&9.26&13.60&12.97\\
		iQuery(Baseline)&10.63&15.96&13.71\\
		DGFNet(Ours)&\textbf{11.25}&\textbf{16.22}&\textbf{14.39}\\
		\bottomrule
	\end{tabular}
\end{table}

\begin{table}
	\caption{Audio-Visual sound separation results on MUSIC21. Best results in \textbf{bold}.}
	\label{tab:music21}
	\begin{tabular}{cccc}
		\toprule
		Methods&SDR$\uparrow$&SIR$\uparrow$&SAR$\uparrow$\\
		\midrule
		The Sound of Pixels&7.52&13.01&11.53\\
		Co-Separation&7.64&13.80&11.30\\
		The Sound of Motions&8.31&\textbf{14.82}&13.11\\
		AVPC-RCoP&9.50&13.74&13.29\\
		iQuery(Baseline)&10.19&14.34&13.79\\
		DGFNet(Ours)&\textbf{10.55}&14.71&\textbf{14.22}\\
		\bottomrule
	\end{tabular}
\end{table}

\begin{table}
	\caption{The results of the ablation study on the impact of different fusion methods.}
	\label{tab:ab}
	\begin{tabular}{cccc}
		\toprule
		Methods&SDR$\uparrow$&SIR$\uparrow$&SAR$\uparrow$\\
		\midrule
		Baseline&10.95&15.56&14.39\\
        +Mul&10.88&15.66&14.06\\
		+DGFM&11.17&15.97&\textbf{14.41}\\
		DGFNet(Ours)&\textbf{11.25}&\textbf{16.22}&14.39\\
		\bottomrule
	\end{tabular}
\end{table}

\section{Conclusion}

This paper presents an end-to-end Audio-Visual Source Separation method based on dynamic gating fusion to address the issue of sound recognition in multi-source environments. The method dynamically adjusts the fusion degree between modal features using a gating mechanism, and integrates an audio attention module to enhance the representational power of audio features, thereby improving the separation performance. Experimental results demonstrate that the proposed method performs excellently in complex background noise or multi-source scenarios, showing strong practicality and promising application prospects.

\section*{Acknowledgements}

This research was financially supported by the National Natural Science Foundation of China (Grant No. 62463029) and the Natural Science Foundation of Xinjiang Uygur Autonomous Region (Grant No. 2015211C288).

\bibliographystyle{ACM-Reference-Format}
\bibliography{sample-base}


\begin{thebibliography}{46}


\ifx \showCODEN    \undefined \def \showCODEN     #1{\unskip}     \fi
\ifx \showDOI      \undefined \def \showDOI       #1{#1}\fi
\ifx \showISBNx    \undefined \def \showISBNx     #1{\unskip}     \fi
\ifx \showISBNxiii \undefined \def \showISBNxiii  #1{\unskip}     \fi
\ifx \showISSN     \undefined \def \showISSN      #1{\unskip}     \fi
\ifx \showLCCN     \undefined \def \showLCCN      #1{\unskip}     \fi
\ifx \shownote     \undefined \def \shownote      #1{#1}          \fi
\ifx \showarticletitle \undefined \def \showarticletitle #1{#1}   \fi
\ifx \showURL      \undefined \def \showURL       {\relax}        \fi
\providecommand\bibfield[2]{#2}
\providecommand\bibinfo[2]{#2}
\providecommand\natexlab[1]{#1}
\providecommand\showeprint[2][]{arXiv:#2}

\bibitem[Afouras et~al\mbox{.}(2020)]%
        {DBLP:conf/eccv/AfourasOCZ20}
\bibfield{author}{\bibinfo{person}{Triantafyllos Afouras}, \bibinfo{person}{Andrew Owens}, \bibinfo{person}{Joon~Son Chung}, {and} \bibinfo{person}{Andrew Zisserman}.} \bibinfo{year}{2020}\natexlab{}.
\newblock \showarticletitle{Self-supervised Learning of Audio-Visual Objects from Video}. In \bibinfo{booktitle}{\emph{Computer Vision - {ECCV} 2020 - 16th European Conference, Glasgow, UK, August 23-28, 2020, Proceedings, Part {XVIII}}} \emph{(\bibinfo{series}{Lecture Notes in Computer Science}, Vol.~\bibinfo{volume}{12363})}. \bibinfo{publisher}{Springer}, \bibinfo{pages}{208--224}.
\newblock
\urldef\tempurl%
\url{https://doi.org/10.1007/978-3-030-58523-5\_13}
\showDOI{\tempurl}


\bibitem[Chen et~al\mbox{.}(2023)]%
        {DBLP:conf/cvpr/ChenZLYZS23}
\bibfield{author}{\bibinfo{person}{Jiaben Chen}, \bibinfo{person}{Renrui Zhang}, \bibinfo{person}{Dongze Lian}, \bibinfo{person}{Jiaqi Yang}, \bibinfo{person}{Ziyao Zeng}, {and} \bibinfo{person}{Jianbo Shi}.} \bibinfo{year}{2023}\natexlab{}.
\newblock \showarticletitle{iQuery: Instruments as Queries for Audio-Visual Sound Separation}. In \bibinfo{booktitle}{\emph{{IEEE/CVF} Conference on Computer Vision and Pattern Recognition, {CVPR} 2023, Vancouver, BC, Canada, June 17-24, 2023}}. \bibinfo{publisher}{{IEEE}}, \bibinfo{pages}{14675--14686}.
\newblock
\urldef\tempurl%
\url{https://doi.org/10.1109/CVPR52729.2023.01410}
\showDOI{\tempurl}


\bibitem[Cheng et~al\mbox{.}(2023)]%
        {DBLP:conf/iclr/ChengLW023}
\bibfield{author}{\bibinfo{person}{Haoyue Cheng}, \bibinfo{person}{Zhaoyang Liu}, \bibinfo{person}{Wayne Wu}, {and} \bibinfo{person}{Limin Wang}.} \bibinfo{year}{2023}\natexlab{}.
\newblock \showarticletitle{Filter-Recovery Network for Multi-Speaker Audio-Visual Speech Separation}. In \bibinfo{booktitle}{\emph{The Eleventh International Conference on Learning Representations, {ICLR} 2023, Kigali, Rwanda, May 1-5, 2023}}. \bibinfo{publisher}{OpenReview.net}.
\newblock
\urldef\tempurl%
\url{https://openreview.net/forum?id=fiB2RjmgwQ6}
\showURL{%
\tempurl}


\bibitem[Cheng et~al\mbox{.}(2020)]%
        {DBLP:conf/mm/ChengWPFZ20}
\bibfield{author}{\bibinfo{person}{Ying Cheng}, \bibinfo{person}{Ruize Wang}, \bibinfo{person}{Zhihao Pan}, \bibinfo{person}{Rui Feng}, {and} \bibinfo{person}{Yuejie Zhang}.} \bibinfo{year}{2020}\natexlab{}.
\newblock \showarticletitle{Look, Listen, and Attend: Co-Attention Network for Self-Supervised Audio-Visual Representation Learning}. In \bibinfo{booktitle}{\emph{{MM} '20: The 28th {ACM} International Conference on Multimedia, Virtual Event / Seattle, WA, USA, October 12-16, 2020}}. \bibinfo{publisher}{{ACM}}, \bibinfo{pages}{3884--3892}.
\newblock
\urldef\tempurl%
\url{https://doi.org/10.1145/3394171.3413869}
\showDOI{\tempurl}


\bibitem[Ding et~al\mbox{.}(2022)]%
        {DBLP:conf/cvpr/DingLYQXCWX22}
\bibfield{author}{\bibinfo{person}{Shuangrui Ding}, \bibinfo{person}{Maomao Li}, \bibinfo{person}{Tianyu Yang}, \bibinfo{person}{Rui Qian}, \bibinfo{person}{Haohang Xu}, \bibinfo{person}{Qingyi Chen}, \bibinfo{person}{Jue Wang}, {and} \bibinfo{person}{Hongkai Xiong}.} \bibinfo{year}{2022}\natexlab{}.
\newblock \showarticletitle{Motion-aware Contrastive Video Representation Learning via Foreground-background Merging}. In \bibinfo{booktitle}{\emph{{IEEE/CVF} Conference on Computer Vision and Pattern Recognition, {CVPR} 2022, New Orleans, LA, USA, June 18-24, 2022}}. \bibinfo{publisher}{{IEEE}}, \bibinfo{pages}{9706--9716}.
\newblock
\urldef\tempurl%
\url{https://doi.org/10.1109/CVPR52688.2022.00949}
\showDOI{\tempurl}


\bibitem[Dosovitskiy et~al\mbox{.}({[n.\,d.]})]%
        {DBLP:conf/iclr/DosovitskiyB0WZ21}
\bibfield{author}{\bibinfo{person}{Alexey Dosovitskiy}, \bibinfo{person}{Lucas Beyer}, \bibinfo{person}{Alexander Kolesnikov}, \bibinfo{person}{Dirk Weissenborn}, \bibinfo{person}{Xiaohua Zhai}, \bibinfo{person}{Thomas Unterthiner}, \bibinfo{person}{Mostafa Dehghani}, \bibinfo{person}{Matthias Minderer}, \bibinfo{person}{Georg Heigold}, \bibinfo{person}{Sylvain Gelly}, \bibinfo{person}{Jakob Uszkoreit}, {and} \bibinfo{person}{Neil Houlsby}.} \bibinfo{year}{[n.\,d.]}\natexlab{}.
\newblock \showarticletitle{An Image is Worth 16x16 Words: Transformers for Image Recognition at Scale}. In \bibinfo{booktitle}{\emph{9th International Conference on Learning Representations, {ICLR} 2021, Virtual Event, Austria, May 3-7, 2021}}. \bibinfo{publisher}{OpenReview.net}.
\newblock
\urldef\tempurl%
\url{https://openreview.net/forum?id=YicbFdNTTy}
\showURL{%
\tempurl}


\bibitem[Ephrat et~al\mbox{.}(2018)]%
        {DBLP:journals/tog/EphratMLDWHFR18}
\bibfield{author}{\bibinfo{person}{Ariel Ephrat}, \bibinfo{person}{Inbar Mosseri}, \bibinfo{person}{Oran Lang}, \bibinfo{person}{Tali Dekel}, \bibinfo{person}{Kevin Wilson}, \bibinfo{person}{Avinatan Hassidim}, \bibinfo{person}{William~T. Freeman}, {and} \bibinfo{person}{Michael Rubinstein}.} \bibinfo{year}{2018}\natexlab{}.
\newblock \showarticletitle{Looking to listen at the cocktail party: a speaker-independent audio-visual model for speech separation}.
\newblock \bibinfo{journal}{\emph{{ACM} Trans. Graph.}} \bibinfo{volume}{37}, \bibinfo{number}{4} (\bibinfo{year}{2018}), \bibinfo{pages}{112}.
\newblock
\urldef\tempurl%
\url{https://doi.org/10.1145/3197517.3201357}
\showDOI{\tempurl}


\bibitem[Gan et~al\mbox{.}(2020)]%
        {DBLP:conf/cvpr/GanHZT020}
\bibfield{author}{\bibinfo{person}{Chuang Gan}, \bibinfo{person}{Deng Huang}, \bibinfo{person}{Hang Zhao}, \bibinfo{person}{Joshua~B. Tenenbaum}, {and} \bibinfo{person}{Antonio Torralba}.} \bibinfo{year}{2020}\natexlab{}.
\newblock \showarticletitle{Music Gesture for Visual Sound Separation}. In \bibinfo{booktitle}{\emph{2020 {IEEE/CVF} Conference on Computer Vision and Pattern Recognition, {CVPR} 2020, Seattle, WA, USA, June 13-19, 2020}}. \bibinfo{publisher}{Computer Vision Foundation / {IEEE}}, \bibinfo{pages}{10475--10484}.
\newblock
\urldef\tempurl%
\url{https://doi.org/10.1109/CVPR42600.2020.01049}
\showDOI{\tempurl}


\bibitem[Gao et~al\mbox{.}(2018)]%
        {DBLP:conf/eccv/GaoFG18}
\bibfield{author}{\bibinfo{person}{Ruohan Gao}, \bibinfo{person}{Rog{\'{e}}rio~Schmidt Feris}, {and} \bibinfo{person}{Kristen Grauman}.} \bibinfo{year}{2018}\natexlab{}.
\newblock \showarticletitle{Learning to Separate Object Sounds by Watching Unlabeled Video}. In \bibinfo{booktitle}{\emph{Computer Vision - {ECCV} 2018 - 15th European Conference, Munich, Germany, September 8-14, 2018, Proceedings, Part {III}}} \emph{(\bibinfo{series}{Lecture Notes in Computer Science}, Vol.~\bibinfo{volume}{11207})}. \bibinfo{publisher}{Springer}, \bibinfo{pages}{36--54}.
\newblock
\urldef\tempurl%
\url{https://doi.org/10.1007/978-3-030-01219-9\_3}
\showDOI{\tempurl}


\bibitem[Gao and Grauman(2019)]%
        {DBLP:conf/iccv/GaoG19}
\bibfield{author}{\bibinfo{person}{Ruohan Gao} {and} \bibinfo{person}{Kristen Grauman}.} \bibinfo{year}{2019}\natexlab{}.
\newblock \showarticletitle{Co-Separating Sounds of Visual Objects}. In \bibinfo{booktitle}{\emph{2019 {IEEE/CVF} International Conference on Computer Vision, {ICCV} 2019, Seoul, Korea (South), October 27 - November 2, 2019}}. \bibinfo{publisher}{{IEEE}}, \bibinfo{pages}{3878--3887}.
\newblock
\urldef\tempurl%
\url{https://doi.org/10.1109/ICCV.2019.00398}
\showDOI{\tempurl}


\bibitem[Gao and Grauman(2021)]%
        {DBLP:conf/cvpr/GaoG21}
\bibfield{author}{\bibinfo{person}{Ruohan Gao} {and} \bibinfo{person}{Kristen Grauman}.} \bibinfo{year}{2021}\natexlab{}.
\newblock \showarticletitle{VisualVoice: Audio-Visual Speech Separation With Cross-Modal Consistency}. In \bibinfo{booktitle}{\emph{{IEEE} Conference on Computer Vision and Pattern Recognition, {CVPR} 2021, virtual, June 19-25, 2021}}. \bibinfo{publisher}{Computer Vision Foundation / {IEEE}}, \bibinfo{pages}{15495--15505}.
\newblock
\urldef\tempurl%
\url{https://doi.org/10.1109/CVPR46437.2021.01524}
\showDOI{\tempurl}


\bibitem[Griffin and Lim(1983)]%
        {DBLP:conf/icassp/GriffinL83}
\bibfield{author}{\bibinfo{person}{Daniel~W. Griffin} {and} \bibinfo{person}{Jae~S. Lim}.} \bibinfo{year}{1983}\natexlab{}.
\newblock \showarticletitle{Signal estimation from modified short-time Fourier transform}. In \bibinfo{booktitle}{\emph{{IEEE} International Conference on Acoustics, Speech, and Signal Processing, {ICASSP} '83, Boston, Massachusetts, USA, April 14-16, 1983}}. \bibinfo{publisher}{{IEEE}}, \bibinfo{pages}{804--807}.
\newblock
\urldef\tempurl%
\url{https://doi.org/10.1109/ICASSP.1983.1172092}
\showDOI{\tempurl}


\bibitem[Haykin and Chen(2005)]%
        {DBLP:journals/neco/HaykinC05}
\bibfield{author}{\bibinfo{person}{Simon Haykin} {and} \bibinfo{person}{Zhe Chen}.} \bibinfo{year}{2005}\natexlab{}.
\newblock \showarticletitle{The Cocktail Party Problem}.
\newblock \bibinfo{journal}{\emph{Neural Comput.}} \bibinfo{volume}{17}, \bibinfo{number}{9} (\bibinfo{year}{2005}), \bibinfo{pages}{1875--1902}.
\newblock
\urldef\tempurl%
\url{https://doi.org/10.1162/0899766054322964}
\showDOI{\tempurl}


\bibitem[Hershey et~al\mbox{.}(2016)]%
        {DBLP:conf/icassp/HersheyCRW16}
\bibfield{author}{\bibinfo{person}{John~R. Hershey}, \bibinfo{person}{Zhuo Chen}, \bibinfo{person}{Jonathan~Le Roux}, {and} \bibinfo{person}{Shinji Watanabe}.} \bibinfo{year}{2016}\natexlab{}.
\newblock \showarticletitle{Deep clustering: Discriminative embeddings for segmentation and separation}. In \bibinfo{booktitle}{\emph{2016 {IEEE} International Conference on Acoustics, Speech and Signal Processing, {ICASSP} 2016, Shanghai, China, March 20-25, 2016}}. \bibinfo{publisher}{{IEEE}}, \bibinfo{pages}{31--35}.
\newblock
\urldef\tempurl%
\url{https://doi.org/10.1109/ICASSP.2016.7471631}
\showDOI{\tempurl}


\bibitem[Hussain(2016)]%
        {Hussain}
\bibfield{author}{\bibinfo{person}{Abrar Hussain}.} \bibinfo{year}{2016}\natexlab{}.
\newblock \showarticletitle{Evaluation of multichannel speech signal separation using Independent Component Analysis}. In \bibinfo{booktitle}{\emph{2016 IEEE Students' Conference on Electrical, Electronics and Computer Science (SCEECS)}}. \bibinfo{pages}{1--7}.
\newblock
\urldef\tempurl%
\url{https://doi.org/10.1109/SCEECS.2016.7509339}
\showDOI{\tempurl}


\bibitem[Islam et~al\mbox{.}(2024)]%
        {DBLP:conf/wacv/IslamNKWYT24}
\bibfield{author}{\bibinfo{person}{Md.~Amirul Islam}, \bibinfo{person}{Seyed~Shahabeddin Nabavi}, \bibinfo{person}{Irina Kezele}, \bibinfo{person}{Yang Wang}, \bibinfo{person}{Yuanhao Yu}, {and} \bibinfo{person}{Jin Tang}.} \bibinfo{year}{2024}\natexlab{}.
\newblock \showarticletitle{Visually Guided Audio Source Separation with Meta Consistency Learning}. In \bibinfo{booktitle}{\emph{{IEEE/CVF} Winter Conference on Applications of Computer Vision, {WACV} 2024, Waikoloa, HI, USA, January 3-8, 2024}}. \bibinfo{publisher}{{IEEE}}, \bibinfo{pages}{3002--3011}.
\newblock
\urldef\tempurl%
\url{https://doi.org/10.1109/WACV57701.2024.00299}
\showDOI{\tempurl}


\bibitem[Ji et~al\mbox{.}(2023)]%
        {DBLP:journals/tmm/JiMXLS23}
\bibfield{author}{\bibinfo{person}{Yanli Ji}, \bibinfo{person}{Shuo Ma}, \bibinfo{person}{Xing Xu}, \bibinfo{person}{Xuelong Li}, {and} \bibinfo{person}{Heng~Tao Shen}.} \bibinfo{year}{2023}\natexlab{}.
\newblock \showarticletitle{Self-Supervised Fine-Grained Cycle-Separation Network {(FSCN)} for Visual-Audio Separation}.
\newblock \bibinfo{journal}{\emph{{IEEE} Trans. Multim.}}  \bibinfo{volume}{25} (\bibinfo{year}{2023}), \bibinfo{pages}{5864--5876}.
\newblock
\urldef\tempurl%
\url{https://doi.org/10.1109/TMM.2022.3200282}
\showDOI{\tempurl}


\bibitem[Kalkhorani et~al\mbox{.}({[n.\,d.]})]%
        {DBLP:conf/icassp/Kalkhorani00XW24}
\bibfield{author}{\bibinfo{person}{Vahid~Ahmadi Kalkhorani}, \bibinfo{person}{Anurag Kumar}, \bibinfo{person}{Ke Tan}, \bibinfo{person}{Buye Xu}, {and} \bibinfo{person}{DeLiang Wang}.} \bibinfo{year}{[n.\,d.]}\natexlab{}.
\newblock \showarticletitle{Audiovisual Speaker Separation with Full- and Sub-Band Modeling in the Time-Frequency Domain}. In \bibinfo{booktitle}{\emph{{IEEE} International Conference on Acoustics, Speech and Signal Processing, {ICASSP} 2024, Seoul, Republic of Korea, April 14-19, 2024}}. \bibinfo{publisher}{{IEEE}}, \bibinfo{pages}{12001--12005}.
\newblock
\urldef\tempurl%
\url{https://doi.org/10.1109/ICASSP48485.2024.10446297}
\showDOI{\tempurl}


\bibitem[Majumder et~al\mbox{.}(2021)]%
        {DBLP:conf/iccv/MajumderAG21}
\bibfield{author}{\bibinfo{person}{Sagnik Majumder}, \bibinfo{person}{Ziad Al{-}Halah}, {and} \bibinfo{person}{Kristen Grauman}.} \bibinfo{year}{2021}\natexlab{}.
\newblock \showarticletitle{Move2Hear: Active Audio-Visual Source Separation}. In \bibinfo{booktitle}{\emph{2021 {IEEE/CVF} International Conference on Computer Vision, {ICCV} 2021, Montreal, QC, Canada, October 10-17, 2021}}. \bibinfo{publisher}{{IEEE}}, \bibinfo{pages}{275--285}.
\newblock
\urldef\tempurl%
\url{https://doi.org/10.1109/ICCV48922.2021.00034}
\showDOI{\tempurl}


\bibitem[Majumder and Grauman(2022)]%
        {DBLP:conf/eccv/MajumderG22}
\bibfield{author}{\bibinfo{person}{Sagnik Majumder} {and} \bibinfo{person}{Kristen Grauman}.} \bibinfo{year}{2022}\natexlab{}.
\newblock \showarticletitle{Active Audio-Visual Separation of Dynamic Sound Sources}. In \bibinfo{booktitle}{\emph{Computer Vision - {ECCV} 2022 - 17th European Conference, Tel Aviv, Israel, October 23-27, 2022, Proceedings, Part {XXXIX}}} \emph{(\bibinfo{series}{Lecture Notes in Computer Science}, Vol.~\bibinfo{volume}{13699})}. \bibinfo{publisher}{Springer}, \bibinfo{pages}{551--569}.
\newblock
\urldef\tempurl%
\url{https://doi.org/10.1007/978-3-031-19842-7\_32}
\showDOI{\tempurl}


\bibitem[Mao et~al\mbox{.}(2021)]%
        {Mao}
\bibfield{author}{\bibinfo{person}{Mingyuan Mao}, \bibinfo{person}{Peng Gao}, \bibinfo{person}{Renrui Zhang}, \bibinfo{person}{Honghui Zheng}, \bibinfo{person}{Teli Ma}, \bibinfo{person}{Peng Yan}, \bibinfo{person}{Errui Ding}, \bibinfo{person}{Baochang Zhang}, {and} \bibinfo{person}{Shumin Han}.} \bibinfo{year}{2021}\natexlab{}.
\newblock \showarticletitle{Dual-stream Network for Visual Recognition}.
\newblock \bibinfo{journal}{\emph{Neural Information Processing Systems,Neural Information Processing Systems}} (\bibinfo{year}{2021}).
\newblock
\urldef\tempurl%
\url{https://proceedings.neurips.cc/paper/2021/hash/d56b9fc4b0f1be8871f5e1c40c0067e7-Abstract.html}
\showURL{%
\tempurl}


\bibitem[Ouyang et~al\mbox{.}(2023)]%
        {DBLP:conf/icassp/OuyangHZLGZH23}
\bibfield{author}{\bibinfo{person}{Daliang Ouyang}, \bibinfo{person}{Su He}, \bibinfo{person}{Guozhong Zhang}, \bibinfo{person}{Mingzhu Luo}, \bibinfo{person}{Huaiyong Guo}, \bibinfo{person}{Jian Zhan}, {and} \bibinfo{person}{Zhijie Huang}.} \bibinfo{year}{2023}\natexlab{}.
\newblock \showarticletitle{Efficient Multi-Scale Attention Module with Cross-Spatial Learning}. In \bibinfo{booktitle}{\emph{{IEEE} International Conference on Acoustics, Speech and Signal Processing {ICASSP} 2023, Rhodes Island, Greece, June 4-10, 2023}}. \bibinfo{publisher}{{IEEE}}, \bibinfo{pages}{1--5}.
\newblock
\urldef\tempurl%
\url{https://doi.org/10.1109/ICASSP49357.2023.10096516}
\showDOI{\tempurl}


\bibitem[Pan et~al\mbox{.}(2023)]%
        {DBLP:conf/asru/PanWMGKHR23}
\bibfield{author}{\bibinfo{person}{Zexu Pan}, \bibinfo{person}{Gordon Wichern}, \bibinfo{person}{Yoshiki Masuyama}, \bibinfo{person}{Fran{\c{c}}ois~G. Germain}, \bibinfo{person}{Sameer Khurana}, \bibinfo{person}{Chiori Hori}, {and} \bibinfo{person}{Jonathan~Le Roux}.} \bibinfo{year}{2023}\natexlab{}.
\newblock \showarticletitle{Scenario-Aware Audio-Visual TF-Gridnet for Target Speech Extraction}. In \bibinfo{booktitle}{\emph{{IEEE} Automatic Speech Recognition and Understanding Workshop, {ASRU} 2023, Taipei, Taiwan, December 16-20, 2023}}. \bibinfo{publisher}{{IEEE}}, \bibinfo{pages}{1--8}.
\newblock
\urldef\tempurl%
\url{https://doi.org/10.1109/ASRU57964.2023.10389618}
\showDOI{\tempurl}


\bibitem[Pegg et~al\mbox{.}(2024)]%
        {pegg2024rtfsnet}
\bibfield{author}{\bibinfo{person}{Samuel Pegg}, \bibinfo{person}{Kai Li}, {and} \bibinfo{person}{Xiaolin Hu}.} \bibinfo{year}{2024}\natexlab{}.
\newblock \showarticletitle{{RTFS}-Net: Recurrent Time-Frequency Modelling for Efficient Audio-Visual Speech Separation}. In \bibinfo{booktitle}{\emph{The Twelfth International Conference on Learning Representations}}.
\newblock
\urldef\tempurl%
\url{https://openreview.net/forum?id=PEuDO2EiDr}
\showURL{%
\tempurl}


\bibitem[Raffel et~al\mbox{.}(2014)]%
        {Raffel_McFee}
\bibfield{author}{\bibinfo{person}{Colin Raffel}, \bibinfo{person}{Brian McFee}, \bibinfo{person}{EricJ. Humphrey}, \bibinfo{person}{Justin Salamon}, \bibinfo{person}{Oriol Nieto}, \bibinfo{person}{Dawen Liang}, {and} \bibinfo{person}{DanielP.W. Ellis}.} \bibinfo{year}{2014}\natexlab{}.
\newblock \showarticletitle{MIR\_EVAL: A Transparent Implementation of Common MIR Metrics.}
\newblock \bibinfo{journal}{\emph{International Symposium/Conference on Music Information Retrieval}} (\bibinfo{year}{2014}).
\newblock


\bibitem[Rahman et~al\mbox{.}(2021)]%
        {DBLP:journals/corr/abs-2110-13412}
\bibfield{author}{\bibinfo{person}{Tanzila Rahman}, \bibinfo{person}{Mengyu Yang}, {and} \bibinfo{person}{Leonid Sigal}.} \bibinfo{year}{2021}\natexlab{}.
\newblock \showarticletitle{TriBERT: Full-body Human-centric Audio-visual Representation Learning for Visual Sound Separation}.
\newblock   \bibinfo{volume}{abs/2110.13412} (\bibinfo{year}{2021}).
\newblock
\urldef\tempurl%
\url{https://arxiv.org/abs/2110.13412}
\showURL{%
\tempurl}


\bibitem[Ren et~al\mbox{.}(2017)]%
        {DBLP:journals/pami/RenHG017}
\bibfield{author}{\bibinfo{person}{Shaoqing Ren}, \bibinfo{person}{Kaiming He}, \bibinfo{person}{Ross~B. Girshick}, {and} \bibinfo{person}{Jian Sun}.} \bibinfo{year}{2017}\natexlab{}.
\newblock \showarticletitle{Faster {R-CNN:} Towards Real-Time Object Detection with Region Proposal Networks}.
\newblock \bibinfo{journal}{\emph{{IEEE} Trans. Pattern Anal. Mach. Intell.}} \bibinfo{volume}{39}, \bibinfo{number}{6} (\bibinfo{year}{2017}), \bibinfo{pages}{1137--1149}.
\newblock
\urldef\tempurl%
\url{https://doi.org/10.1109/TPAMI.2016.2577031}
\showDOI{\tempurl}


\bibitem[Ronneberger et~al\mbox{.}(2015)]%
        {DBLP:conf/miccai/RonnebergerFB15}
\bibfield{author}{\bibinfo{person}{Olaf Ronneberger}, \bibinfo{person}{Philipp Fischer}, {and} \bibinfo{person}{Thomas Brox}.} \bibinfo{year}{2015}\natexlab{}.
\newblock \showarticletitle{U-Net: Convolutional Networks for Biomedical Image Segmentation}. In \bibinfo{booktitle}{\emph{Medical Image Computing and Computer-Assisted Intervention - {MICCAI} 2015 - 18th International Conference Munich, Germany, October 5 - 9, 2015, Proceedings, Part {III}}} \emph{(\bibinfo{series}{Lecture Notes in Computer Science}, Vol.~\bibinfo{volume}{9351})}, \bibfield{editor}{\bibinfo{person}{Nassir Navab}, \bibinfo{person}{Joachim Hornegger}, \bibinfo{person}{William M.~Wells III}, {and} \bibinfo{person}{Alejandro~F. Frangi}} (Eds.). \bibinfo{publisher}{Springer}, \bibinfo{pages}{234--241}.
\newblock
\urldef\tempurl%
\url{https://doi.org/10.1007/978-3-319-24574-4\_28}
\showDOI{\tempurl}


\bibitem[Song and Zhang(2024)]%
        {10180219}
\bibfield{author}{\bibinfo{person}{Zengjie Song} {and} \bibinfo{person}{Zhaoxiang Zhang}.} \bibinfo{year}{2024}\natexlab{}.
\newblock \showarticletitle{Visually Guided Sound Source Separation With Audio-Visual Predictive Coding}.
\newblock \bibinfo{journal}{\emph{IEEE Transactions on Neural Networks and Learning Systems}} \bibinfo{volume}{35}, \bibinfo{number}{11} (\bibinfo{year}{2024}), \bibinfo{pages}{15528--15542}.
\newblock
\urldef\tempurl%
\url{https://doi.org/10.1109/TNNLS.2023.3288022}
\showDOI{\tempurl}


\bibitem[Tian et~al\mbox{.}(2021)]%
        {DBLP:conf/cvpr/TianHX21}
\bibfield{author}{\bibinfo{person}{Yapeng Tian}, \bibinfo{person}{Di Hu}, {and} \bibinfo{person}{Chenliang Xu}.} \bibinfo{year}{2021}\natexlab{}.
\newblock \showarticletitle{Cyclic Co-Learning of Sounding Object Visual Grounding and Sound Separation}. In \bibinfo{booktitle}{\emph{{IEEE} Conference on Computer Vision and Pattern Recognition, {CVPR} 2021, virtual, June 19-25, 2021}}. \bibinfo{publisher}{Computer Vision Foundation / {IEEE}}, \bibinfo{pages}{2745--2754}.
\newblock
\urldef\tempurl%
\url{https://doi.org/10.1109/CVPR46437.2021.00277}
\showDOI{\tempurl}


\bibitem[Tzinis et~al\mbox{.}(2021)]%
        {Efthymios}
\bibfield{author}{\bibinfo{person}{Efthymios Tzinis}, \bibinfo{person}{Scott Wisdom}, \bibinfo{person}{Aren Jansen}, \bibinfo{person}{Shawn Hershey}, \bibinfo{person}{Tal Remez}, \bibinfo{person}{Dan Ellis}, {and} \bibinfo{person}{John~R. Hershey}.} \bibinfo{year}{2021}\natexlab{}.
\newblock \showarticletitle{Into the Wild with AudioScope: Unsupervised Audio-Visual Separation of On-Screen Sounds}. In \bibinfo{booktitle}{\emph{9th International Conference on Learning Representations, {ICLR} 2021, Virtual Event, Austria, May 3-7, 2021}}. \bibinfo{publisher}{OpenReview.net}.
\newblock
\urldef\tempurl%
\url{https://openreview.net/forum?id=MDsQkFP1Aw}
\showURL{%
\tempurl}


\bibitem[Tzinis et~al\mbox{.}(2022)]%
        {DBLP:conf/eccv/TzinisWRH22}
\bibfield{author}{\bibinfo{person}{Efthymios Tzinis}, \bibinfo{person}{Scott Wisdom}, \bibinfo{person}{Tal Remez}, {and} \bibinfo{person}{John~R. Hershey}.} \bibinfo{year}{2022}\natexlab{}.
\newblock \showarticletitle{AudioScopeV2: Audio-Visual Attention Architectures for Calibrated Open-Domain On-Screen Sound Separation}. In \bibinfo{booktitle}{\emph{Computer Vision - {ECCV} 2022 - 17th European Conference, Tel Aviv, Israel, October 23-27, 2022, Proceedings, Part {XXXVII}}} \emph{(\bibinfo{series}{Lecture Notes in Computer Science}, Vol.~\bibinfo{volume}{13697})}, \bibfield{editor}{\bibinfo{person}{Shai Avidan}, \bibinfo{person}{Gabriel~J. Brostow}, \bibinfo{person}{Moustapha Ciss{\'{e}}}, \bibinfo{person}{Giovanni~Maria Farinella}, {and} \bibinfo{person}{Tal Hassner}} (Eds.). \bibinfo{publisher}{Springer}, \bibinfo{pages}{368--385}.
\newblock
\urldef\tempurl%
\url{https://doi.org/10.1007/978-3-031-19836-6\_21}
\showDOI{\tempurl}


\bibitem[Vaswani et~al\mbox{.}(2017)]%
        {Vaswani}
\bibfield{author}{\bibinfo{person}{Ashish Vaswani}, \bibinfo{person}{Noam Shazeer}, \bibinfo{person}{Niki Parmar}, \bibinfo{person}{Jakob Uszkoreit}, \bibinfo{person}{Llion Jones}, \bibinfo{person}{AidanN. Gomez}, \bibinfo{person}{Lukasz Kaiser}, {and} \bibinfo{person}{Illia Polosukhin}.} \bibinfo{year}{2017}\natexlab{}.
\newblock \showarticletitle{Attention is All you Need}.
\newblock \bibinfo{journal}{\emph{Neural Information Processing Systems,Neural Information Processing Systems}} (\bibinfo{date}{Jun} \bibinfo{year}{2017}).
\newblock
\urldef\tempurl%
\url{https://proceedings.neurips.cc/paper/2017/hash/3f5ee243547dee91fbd053c1c4a845aa-Abstract.html}
\showURL{%
\tempurl}


\bibitem[Ye et~al\mbox{.}(2024)]%
        {DBLP:conf/wacv/YeYT24}
\bibfield{author}{\bibinfo{person}{Yuxin Ye}, \bibinfo{person}{Wenming Yang}, {and} \bibinfo{person}{Yapeng Tian}.} \bibinfo{year}{2024}\natexlab{}.
\newblock \showarticletitle{{LAVSS:} Location-Guided Audio-Visual Spatial Audio Separation}. In \bibinfo{booktitle}{\emph{{IEEE/CVF} Winter Conference on Applications of Computer Vision, {WACV} 2024, Waikoloa, HI, USA, January 3-8, 2024}}. \bibinfo{publisher}{{IEEE}}, \bibinfo{pages}{5496--5507}.
\newblock
\urldef\tempurl%
\url{https://doi.org/10.1109/WACV57701.2024.00542}
\showDOI{\tempurl}


\bibitem[Yu et~al\mbox{.}(2017)]%
        {DBLP:conf/icassp/YuKT017}
\bibfield{author}{\bibinfo{person}{Dong Yu}, \bibinfo{person}{Morten Kolb{\ae}k}, \bibinfo{person}{Zheng{-}Hua Tan}, {and} \bibinfo{person}{Jesper Jensen}.} \bibinfo{year}{2017}\natexlab{}.
\newblock \showarticletitle{Permutation invariant training of deep models for speaker-independent multi-talker speech separation}. In \bibinfo{booktitle}{\emph{2017 {IEEE} International Conference on Acoustics, Speech and Signal Processing, {ICASSP} 2017, New Orleans, LA, USA, March 5-9, 2017}}. \bibinfo{publisher}{{IEEE}}, \bibinfo{pages}{241--245}.
\newblock
\urldef\tempurl%
\url{https://doi.org/10.1109/ICASSP.2017.7952154}
\showDOI{\tempurl}


\bibitem[Yu et~al\mbox{.}(2022a)]%
        {fsaavn}
\bibfield{author}{\bibinfo{person}{Yinfeng Yu}, \bibinfo{person}{Lele Cao}, \bibinfo{person}{Fuchun Sun}, \bibinfo{person}{Xiaohong Liu}, {and} \bibinfo{person}{Liejun Wang}.} \bibinfo{year}{2022}\natexlab{a}.
\newblock \showarticletitle{Pay Self-Attention to Audio-Visual Navigation}. In \bibinfo{booktitle}{\emph{33rd British Machine Vision Conference 2022, {BMVC} 2022, London, UK, November 21-24, 2022}}. \bibinfo{publisher}{{BMVA} Press}, \bibinfo{pages}{46}.
\newblock


\bibitem[Yu et~al\mbox{.}(2023a)]%
        {ttt}
\bibfield{author}{\bibinfo{person}{Yinfeng Yu}, \bibinfo{person}{Lele Cao}, \bibinfo{person}{Fuchun Sun}, \bibinfo{person}{Chao Yang}, \bibinfo{person}{Huicheng Lai}, {and} \bibinfo{person}{Wenbing Huang}.} \bibinfo{year}{2023}\natexlab{a}.
\newblock \showarticletitle{Echo-Enhanced Embodied Visual Navigation}.
\newblock \bibinfo{journal}{\emph{Neural Computation}} \bibinfo{volume}{35}, \bibinfo{number}{5} (\bibinfo{date}{04} \bibinfo{year}{2023}), \bibinfo{pages}{958--976}.
\newblock
\showISSN{0899-7667}


\bibitem[Yu et~al\mbox{.}(2023b)]%
        {YinfengIJCAI2023MACMA}
\bibfield{author}{\bibinfo{person}{Yinfeng Yu}, \bibinfo{person}{Changan Chen}, \bibinfo{person}{Lele Cao}, \bibinfo{person}{Fangkai Yang}, \bibinfo{person}{Wenbing Huang}, {and} \bibinfo{person}{Fuchun Sun}.} \bibinfo{year}{2023}\natexlab{b}.
\newblock \showarticletitle{Measuring Acoustics with Collaborative Multiple Agents}. In \bibinfo{booktitle}{\emph{The 32nd International Joint Conference on Artificial Intelligence, {IJCAI} 2023, Macao, 19th-25th August 2023}}.
\newblock


\bibitem[Yu et~al\mbox{.}(2022b)]%
        {SAAVN}
\bibfield{author}{\bibinfo{person}{Yinfeng Yu}, \bibinfo{person}{Wenbing Huang}, \bibinfo{person}{Fuchun Sun}, \bibinfo{person}{Changan Chen}, \bibinfo{person}{Yikai Wang}, {and} \bibinfo{person}{Xiaohong Liu}.} \bibinfo{year}{2022}\natexlab{b}.
\newblock \showarticletitle{Sound Adversarial Audio-Visual Navigation}. In \bibinfo{booktitle}{\emph{The Tenth International Conference on Learning Representations, {ICLR} 2022, Virtual Event, April 25-29, 2022}}. \bibinfo{publisher}{OpenReview.net}.
\newblock


\bibitem[Yuan et~al\mbox{.}({[n.\,d.]})]%
        {Yuan}
\bibfield{author}{\bibinfo{person}{Li Yuan}, \bibinfo{person}{Yunpeng Chen}, \bibinfo{person}{Tao Wang}, \bibinfo{person}{Weihao Yu}, \bibinfo{person}{Yujun Shi}, \bibinfo{person}{Zihang Jiang}, \bibinfo{person}{Francis E.~H. Tay}, \bibinfo{person}{Jiashi Feng}, {and} \bibinfo{person}{Shuicheng Yan}.} \bibinfo{year}{[n.\,d.]}\natexlab{}.
\newblock \showarticletitle{Tokens-to-Token ViT: Training Vision Transformers from Scratch on ImageNet}. In \bibinfo{booktitle}{\emph{2021 IEEE/CVF International Conference on Computer Vision (ICCV)}}.
\newblock
\urldef\tempurl%
\url{https://doi.org/10.1109/ICCV48922.2021.00060}
\showDOI{\tempurl}


\bibitem[Zhao et~al\mbox{.}(2017)]%
        {Zhao_Gallo_Frosio_Kautz_2017}
\bibfield{author}{\bibinfo{person}{Hang Zhao}, \bibinfo{person}{Orazio Gallo}, \bibinfo{person}{Iuri Frosio}, {and} \bibinfo{person}{Jan Kautz}.} \bibinfo{year}{2017}\natexlab{}.
\newblock \showarticletitle{Loss Functions for Image Restoration With Neural Networks}.
\newblock \bibinfo{journal}{\emph{IEEE Transactions on Computational Imaging}} (\bibinfo{year}{2017}), \bibinfo{pages}{47–57}.
\newblock
\urldef\tempurl%
\url{https://doi.org/10.1109/TCI.2016.2644865}
\showDOI{\tempurl}


\bibitem[Zhao et~al\mbox{.}(2019)]%
        {DBLP:conf/iccv/ZhaoGM019}
\bibfield{author}{\bibinfo{person}{Hang Zhao}, \bibinfo{person}{Chuang Gan}, \bibinfo{person}{Wei{-}Chiu Ma}, {and} \bibinfo{person}{Antonio Torralba}.} \bibinfo{year}{2019}\natexlab{}.
\newblock \showarticletitle{The Sound of Motions}. In \bibinfo{booktitle}{\emph{2019 {IEEE/CVF} International Conference on Computer Vision, {ICCV} 2019, Seoul, Korea (South), October 27 - November 2, 2019}}. \bibinfo{publisher}{{IEEE}}, \bibinfo{pages}{1735--1744}.
\newblock
\urldef\tempurl%
\url{https://doi.org/10.1109/ICCV.2019.00182}
\showDOI{\tempurl}


\bibitem[Zhao et~al\mbox{.}(2018)]%
        {Zhao_2018_ECCV}
\bibfield{author}{\bibinfo{person}{Hang Zhao}, \bibinfo{person}{Chuang Gan}, \bibinfo{person}{Andrew Rouditchenko}, \bibinfo{person}{Carl Vondrick}, \bibinfo{person}{Josh McDermott}, {and} \bibinfo{person}{Antonio Torralba}.} \bibinfo{year}{2018}\natexlab{}.
\newblock \showarticletitle{The Sound of Pixels}. In \bibinfo{booktitle}{\emph{The European Conference on Computer Vision (ECCV)}}.
\newblock
\urldef\tempurl%
\url{http://arxiv.org/abs/1804.03160}
\showURL{%
\tempurl}


\bibitem[Zhou et~al\mbox{.}(2022)]%
        {zhou2022detecting}
\bibfield{author}{\bibinfo{person}{Xingyi Zhou}, \bibinfo{person}{Rohit Girdhar}, \bibinfo{person}{Armand Joulin}, \bibinfo{person}{Philipp Kr{\"a}henb{\"u}hl}, {and} \bibinfo{person}{Ishan Misra}.} \bibinfo{year}{2022}\natexlab{}.
\newblock \showarticletitle{Detecting Twenty-thousand Classes using Image-level Supervision}. In \bibinfo{booktitle}{\emph{ECCV}} \emph{(\bibinfo{series}{Lecture Notes in Computer Science}, Vol.~\bibinfo{volume}{13669})}. \bibinfo{publisher}{Springer}, \bibinfo{pages}{350--368}.
\newblock
\urldef\tempurl%
\url{https://doi.org/10.1007/978-3-031-20077-9\_21}
\showDOI{\tempurl}


\bibitem[Zhu and Rahtu(2020)]%
        {DBLP:conf/accv/0001R20}
\bibfield{author}{\bibinfo{person}{Lingyu Zhu} {and} \bibinfo{person}{Esa Rahtu}.} \bibinfo{year}{2020}\natexlab{}.
\newblock \showarticletitle{Visually Guided Sound Source Separation Using Cascaded Opponent Filter Network}. In \bibinfo{booktitle}{\emph{Computer Vision - {ACCV} 2020 - 15th Asian Conference on Computer Vision, Kyoto, Japan, November 30 - December 4, 2020, Revised Selected Papers, Part {VI}}} \emph{(\bibinfo{series}{Lecture Notes in Computer Science}, Vol.~\bibinfo{volume}{12627})}, \bibfield{editor}{\bibinfo{person}{Hiroshi Ishikawa}, \bibinfo{person}{Cheng{-}Lin Liu}, \bibinfo{person}{Tom{\'{a}}s Pajdla}, {and} \bibinfo{person}{Jianbo Shi}} (Eds.). \bibinfo{publisher}{Springer}, \bibinfo{pages}{409--426}.
\newblock
\urldef\tempurl%
\url{https://doi.org/10.1007/978-3-030-69544-6\_25}
\showDOI{\tempurl}


\bibitem[Zhu and Rahtu(2022)]%
        {Zhu_Rahtu_2022}
\bibfield{author}{\bibinfo{person}{Lingyu Zhu} {and} \bibinfo{person}{Esa Rahtu}.} \bibinfo{year}{2022}\natexlab{}.
\newblock \showarticletitle{Visually Guided Sound Source Separation and Localization using Self-Supervised Motion Representations}. In \bibinfo{booktitle}{\emph{2022 IEEE/CVF Winter Conference on Applications of Computer Vision (WACV)}}. \bibinfo{pages}{2171–2181}.
\newblock
\urldef\tempurl%
\url{https://doi.org/10.1109/wacv51458.2022.00223}
\showDOI{\tempurl}


\end{thebibliography}

\end{document}